# Radio-Frequency Multiply-And-Accumulate Operations with Spintronic Synapses

Nathan Leroux[1*], Danijela Marković[1], Erwann Martin[2], Teodora Petrisor[2], Damien Querlioz[3], Alice Mizrahi[1] and Julie Grollier[1]

[1] - Unité Mixte de Physique, CNRS, Thales, Université Paris-Saclay, 91767 Palaiseau, France
[2] - Thales Research and Technology, 91767 Palaiseau, France
[3] - Université Paris-Saclay, CNRS, Centre de Nanosciences et de Nanotechnologies, 91120 Palaiseau, France

* nathan.leroux@cnrs-thales.fr

Exploiting the physics of nanoelectronic devices is a major lead for implementing compact, fast, and energy efficient artificial intelligence. In this work, we propose an original road in this direction, where assemblies of spintronic resonators used as artificial synapses can classify analogue radio-frequency signals directly without digitalization. The resonators convert the radio-frequency input signals into direct voltages through the spin-diode effect. In the process, they multiply the input signals by a synaptic weight, which depends on their resonance frequency. We demonstrate through physical simulations with parameters extracted from experimental devices that frequency-multiplexed assemblies of resonators implement the cornerstone operation of artificial neural networks, the Multiply-And-Accumulate (MAC), directly on microwave inputs. The results show that even with a non-ideal realistic model, the outputs obtained with our architecture remain comparable to that of a traditional MAC operation. Using a conventional machine learning framework augmented with equations describing the physics of spintronic resonators, we train a single layer neural network to classify radio-frequency signals encoding 8x8 pixel handwritten digits pictures. The spintronic neural network recognizes the digits with an accuracy of 99.96 %, equivalent to purely software neural networks. This MAC implementation offers a promising solution for fast, low-power radio-frequency classification applications, and another building block for spintronic deep neural networks.

## I. Introduction

Radio-frequency (RF) signals are ubiquitous today [1]. Finding ways to automatically recognize and classify these signals is important for numerous applications such as medicine [2–4], RF fingerprinting [5], gesture sensing [6], radar applications [7], aerial vehicle detection and identification [8]. Artificial neural networks have proven to be more accurate and more resilient to real-world conditions (noisy electromagnetic environment, imperfect RF components or antennas etc.) than more conventional algorithms that rely on specific features extractors and complex analysis tools [1]. Currently, applying artificial neural networks to RF signals requires to first digitize the signal sensed by the antenna and then to use and run a neural network on conventional CMOS-based hardware (such as a central processing unit, graphics processingunit (GPU), field-programmable gate array, or application specific integrated circuit). Both stages of the process are computationally heavy, leading to delays (a few miliseconds) and high power and energy consumption (hundreds of Watts) [9]. To decrease the size and the dependency to cloud computing of embedded RF devices, it is thus essential to build fast and low power systems that integrate both RF signal analyzers and in situ Artificial Intelligence accelerators.

Presently, the most promising Artificial Intelligence algorithms are based on deep neural networks [10], which contain several layers of artificial neurons, each of them linked by synaptic connections: in each layer of an artificial neural network, the neuron signals are multiplied by synaptic weights, summed and injected into a neuron of the following layer (see Fig. 1a). This elementary operation is called Multiply-And-Accumulate (MAC). In a computer using the von Neumann architecture, weight multiplications and sums are performed by processing units, whereas synaptic weight values are stored in spatially separated memory units. In such architecture, the data flow between the processing and memory units induces a slowdown and excess energy consumption [11] that can be avoided by implementing the MAC operation in hardware, using in situ memory devices emulating neurons and synapses [12–17]. Neurons that take dc inputs and convert them to microwave signals have been demonstrated using spintronic nano-oscillators [18–23] and CMOS ring oscillators [24,25]. However, to this day there is no demonstration of tunable artificial synapses that directly perform MAC operations on microwave signals.

In this work, we show through theoretical analysis and numerical simulations that spintronic resonators, which are devices similar to spintronic oscillators, may be used as RF nano-synapses. These resonators apply synaptic weights to microwave signals through the spin-diode effect [26–28]: because of magnetic resonance, when an RF current passes through a spintronic resonator, the resistance of the resonator is forced to oscillate at the same frequency as the RF current, thus rectifying the input power. The amplitude of the rectified voltage depends on the difference between the frequency of the RF signal and the frequency of the device resonance [26–29]. We can thus use spintronic resonators as artificial synapses that take microwave power as input, rectify direct voltage as output and store synaptic weights encoded in their resonance frequency.

We first explain the RF neural network architecture with inputs encoded in microwave powers and show how frequency multiplexing can be used to make a simple and compact implementation of the MAC operation. We explain how the spintronic resonators can implement tunable synaptic weights through the spin-diode effect. To be realistic, we have to consider the non-linear behavior of the spintronic resonators, that, may in principle represent a challenge. Physical simulations using parameters extracted from experimental devices demonstrate that our architecture is equivalent to a MAC operation even when the resonator non-linearities are considered. Furthermore, the response of a resonator under the superimposition of multiple RF signals with different frequencies is non-trivial. We demonstrate through dynamical simulations that we can simulate this response with a simple analytical model. This allows us to highlight the fundamental requirements for this RF MAC proposal. Finally, we show through simulations of a complete network that a single layer of our RF neural network can be used to classify microwave signals encoding handwritten digits images, with an accuracy equivalent to software MACs.

## II. Principle of resonator-based MAC operations on radio-frequency signals: study in idealized conditions

As represented Fig. 1.a, the Multiply-And-Accumulate operation is a weighted sum of N input values. In our proposal, these N input values are encoded in the microwave powers $P_i$ of N RF signals of index $i$, each with a different frequency. A MAC operation is performed by sending these N RF signals simultaneously to a chain of N resonators, indexed by $k$, wired in series (Fig.1.b). A neural network with M outputs requires M different resonator chains, indexed by $j$. The goal of this section is to show that the voltage across each chain $j$ can be seen as a weighted sum of the input microwave powers $P_i$:

$$U_j = \sum_{i=0}^{N-1} P_i W_{ji}, \qquad (1)$$

where $W_{ji}$ is a synaptic weight between the input $i$ and the output $j$, determined by the physics of the spin diode effect. All resonators in a chain $j$ can contribute to $W_{ji}$, depending on their frequency distribution and the frequency of the input signals.

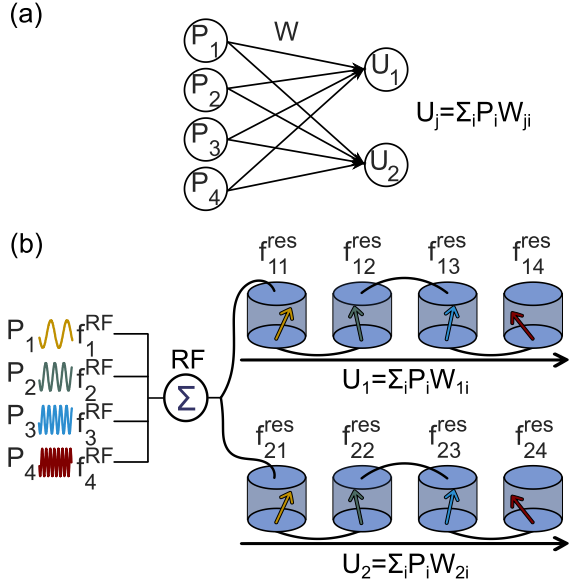

**Fig. 1** *(a) Multiply-And-Accumulate operation: the neural signals $P_1$, $P_2$, $P_3$ and $P_4$ are multiplied by different synaptic weights $W_{ji}$ and summed. (b) Multiply-And-Accumulate operation with different radio-frequency signals sent simultaneously in 2 chains of resonators: each resonator rectifies mostly one of the input signals, hence multiplying it by a weight. The chain voltages are the sum of all their resonators voltages.*

Following the universal, and experimentally validated [30,31] auto-oscillator model described in the paper of A. Slavin and V. Tiberkevich [29], the rectification voltage of an ideal spintronic resonator $k$ in chain $j$ submitted to an RF power $P_i$ with angular frequency $\omega_i^{RF}$ due to the spin-diode effect, displayed in Fig. 2.a. is:

$$\begin{cases} v_{kji} = P_i G(\Delta f_{kji}) = P_i \dfrac{\omega_i^{RF} - \omega_{kj}^{res}}{\Gamma_{kj}^{res\,2} + \left(\omega_i^{RF} - \omega_{kj}^{res}\right)^2} \beta \\ \omega_{kj}^{res} = 2\pi f_{kj}^{res} \\ \Gamma_{kj}^{res} = \alpha \omega_{kj}^{res} \end{cases} \quad (2)$$

where $\omega_{kj}^{res}$ is the angular frequency of resonance, $\Gamma_{kj}^{res}$ is the resonance linewidth, α is the magnetic damping, $\Delta f_{kji}$ is the frequency mismatch $\Delta f_{kji} = f_i^{RF} - f_{kj}^{res}$ and β is a factor that depends on several characteristics of the resonators, for instance the magnetoresistance if they are magnetic tunnel junctions [18].

Because there are often different torques acting on the magnetization dynamics, the spin-diode effect usually leads to the sum of a symmetric and an anti-symmetric part in the rectified voltage [27]. For instance, in Magnetic Tunnel Junctions, the Field-Like torque usually leads to an anti-symmetric component and the Spin-Transfer-Torque usually leads to a symmetric part. For clarity, we chose to focus on the anti-symmetric part in this paper. Results of simulations including the symmetric part are presented in section V.

Each resonator receives simultaneously the N RF input signals. We show in section IV that the resulting rectified voltage is the sum of the dc voltages they generate when they receive each RF signal individually. Therefore, the rectified voltage across each chain is:

$$U_j = \sum_{k=0}^{N-1} \sum_{i=0}^{N-1} v_{kji}. \tag{3}$$

In this work, we propose to wire the resonators of a chain in a head-to-tail configuration, as depicted in Fig. 1b, to cancel the voltage offsets at low frequency ( $\omega_i^{RF} \to 0$ in Eq. 2). Indeed, the RF signal of index $i$ is rectified into a positive voltage by all the resonators of index $k>i$. All these offset voltages would accumulate and become larger than the resonant signals useful for synaptic operations. But if we wire the resonators head-to-tail, then the RF signal of index $i$ is rectified into a positive voltage by all the resonators of even indexes $k>i$, and into a negative voltage by all the resonators of odd indexes $k>i$. Hence the offset generated by the RF signal of index $i$ is approximately compensated, and it remains mostly the voltage rectified by the resonator of index $k=i$. The combination of Eqs. 2 and 3 therefore gives:

$$U_j = \sum_{k=0}^{N-1} \sum_{i=0}^{N-1} P_i G(\Delta f_{kji})(-1)^k \tag{4}$$

where the factor $(-1)^k$ accounts for the head-to-tail wiring. This naturally leads to Eq. 1, with the synaptic weights equal to

$$W_{ji} = \sum_{k=0}^{N-1} G(\Delta f_{kji})(-1)^k. \tag{5}$$

Spintronic resonators are frequency selective. As can be seen in Fig. 2.a and Eq. 2, the rectification voltage drops to zero when $\omega_i^{RF}$ tends toward infinity and to a small offset ($2\alpha$ times smaller than the maximum voltage) when $\omega_i^{RF}$ tends toward 0. For operating a resonator chain as a useful neural network, each resonator in a synaptic chain should be chosen to have a resonance frequency matching the frequency of one of the input signals, so that this resonator features a greater rectification effect on this matching signal. For instance, in Fig. 1.b, the resonator with resonance frequency $f_{12}^{res}$ receives the four RF signals but rectifies most effectively the signal with frequency $f_2^{RF}$. When using the synaptic chain in this configuration, each synaptic weight can be approximated, leading to a simplified expression of Eq. 5: $W_{ji} = G(\Delta f_{iji})(-1)^i$.

This simplified equation highlights that it is possible to tune each synaptic weight $W_{ji}$ by tuning the resonance frequency of the resonator indexed by $k = i$, which plays the role of a synaptic connection between input $i$ and output $j$ (see Fig. 1a). M. Zahedinejad et al [32] demonstrated a voltage gate memristive control of the perpendicular magnetic anisotropy at a ferromagnetic-oxyde interface leading to non-volatile control of the oscillating properties of a spin Hall nano-oscillator. Such a memristive control of the magnetic properties of a spintronic resonator could allow tuning the resonance frequencies in future experimental implementations. To send the sum of the RF signals into the different chains of resonators like in Fig. 1.b, it is possible to build micrometer scale RF combiners with low power dissipation as it was done in [24].

### III. Multiply-And-Accumulate simulations results incorporating device non-linearities

We now quantify the accuracy of the spintronic resonator-based MAC operation compared to an ideal one. Spintronic resonators have an intrinsic non-linear dependence of their frequency and linewidth on the magnetization oscillation amplitude, which is typically expressed as [29]

$$\begin{cases} f^{res}(p) = f^{res}(0)(1 + Np) \\ \Gamma^{res}(p) = 2\pi f^{res}(0)(1 + Qp) \end{cases} \tag{6}$$

where N and Q are nonlinear parameters, and $p$ is the normalized magnetization oscillation power, equal to the square of the magnetization oscillation amplitude. According to [29], $p$ can be expressed as:

$$p = \frac{P^{RF}}{\Gamma^{res}(p)^2 + \left(\omega^{RF} - \omega^{res}(p)\right)^2} \gamma^2 \qquad (7)$$

where $\gamma$ is a proportionality factor between the amplitude of the RF signal and the amplitude of the torque acting on the resonator magnetization. This means that, in real devices, the dependence of $v_{kji}$ with the input power $P^{RF} = P_i$ in Eq. 2 is not perfectly linear, as $\omega_{kj}^{res}$ and $\Gamma_{kj}^{res}$ both depend on $P_i$. In other words, the weights $W_{ji}$ depend on the inputs, which does not correspond to the usual mathematical description of neural networks. This effect, called weight non-linearity, is an issue for learning in hardware neural networks utilizing nanodevices such as memristors to implement MAC operations through physical phenomena.

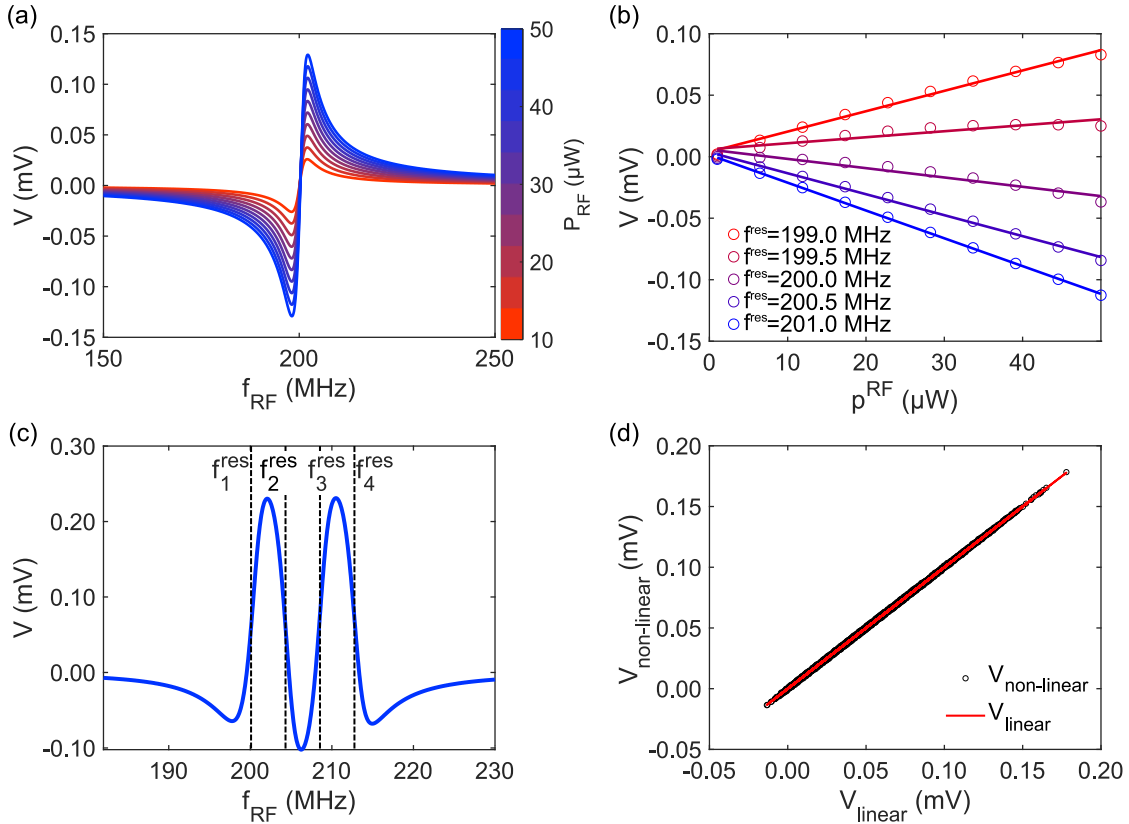

*Fig.2) Analytical calculations based on the realistic spin-diode model described by Eqs. 2 and 6. **(a)** Spin-diode rectified voltage of a spintronic resonator with resonance frequency $f^{res} = 200$ MHz versus the frequency of an input radio-frequency signal for microwave powers between 10 and 50 µW (in colorscale). **(b)** Circles are the spin-diode voltage of a spintronic resonator with different resonance frequencies versus the microwave power of RF signal at a 200 MHz. Solid lines are linear fits. **(c)** Spin-diode voltage of a chain of four spintronic resonators wired head-to-tail with resonance frequencies $f^{res} = 200$ Hz, 204.0 MHz, 208.2 MHz and 212.4 MHz, versus the frequency of a radio-frequency signal of power 50 µW. **(d)** Calculation for the same chain with four different radio-frequency signals for 6561 different combinations of microwave powers for the radio-frequency signals (5 µW, 10 µW, and 15 µW) and different resonance frequencies for the resonators. The scatter dots are the voltages of the calculations with non-linear resonators plotted against the voltages of the calculations with ideal linear resonators. The red solid line corresponds to the calculated voltages with ideal linear resonators plotted against themselves. The root-mean-square deviation between the scatter dots and the red solid line is 0.58 µV and the correlation is 99.98 %.*

To quantify this effect, we choose parameters extracted from experiments on similar structures as the resonators. We take nonlinear coefficients N=0.1 and Q=1 close to values determined experimentally in studies of spin-torque nano-oscillators [29,33–35]. We chose frequencies close to 200 MHz, parameters $\beta = 1.7 \times 10^6$ C$^{-1}$ and $\gamma = 7.1 \times 10^7$ Hz.W$^{-1/2}$ according to experimental values from prior work [33]. Finally, we took a damping parameter $\alpha = 0.01$ corresponding to Permalloy. It is important to note that these parameters are only used as a guideline, as we could also use spin-orbit torque or Oersted field instead of spin-transfer torque, and the materials and the geometry of the spintronic resonators could also be different. These different devices would follow the same model but with different parameters. It is even possible to reduce the nonlinear parameters N and Q using specific fabrication process [34] or geometry [36].

We plot in Fig. 2.b the dependence of the spin-diode voltage of a single resonator as a function of the microwave power of a RF signal, for different resonance frequency values using the analytical model of Eq. 2-8. We see that despite the nonlinearities N and Q, the voltage response can be fitted linearly with the microwave power, indicating that the dependence of the corresponding weight on the input power remains small. Fig. 2.b also shows that the slope can be controlled by tuning the resonance frequency of each resonator. This result confirms that, with realistic devices, the synaptic weights of a resonator-based neural network can be tuned by changing the resonance frequencies of the resonators.

We now compare the resonator-based MAC operation to a perfectly linear MAC operation. We consider four input RF signals of frequencies $f^{RF}$ = 200.0 MHz, 204.0 MHz, 208.2 MHz and 212.4 MHz and simulate a chain of four different resonators as illustrated in Fig. 2.c, with $N_{powers} = 3$ different input powers (5 µW, 10 µW and 15 µW) for each RF signal and $N_{frequencies} = 3$ different resonance frequencies for each resonator, resulting in a set of $N_{powers}^{N^{RF}} \times N_{frequencies}^{N^{res}} = 6561$ different combinations.

This result gives us the performance of the non-linear MAC operation. We then need to define a reference, ideal MAC operation to evaluate the results. For this purpose, for every point, we compute the magnetization oscillation powers of the four diodes with Eq. 7, and store the maximum for each resonator and for each RF signal: $p_{ki}^{max} = max_{realMAC}(p_{ki})$. These maximum oscillation power values serve as a reference to simulate a MAC operation with a chain of four ideal linear spintronic resonators whose resonance frequency and linewidth does not depend on the input RF power. This approach gives the following linear reference model for the MAC operation:

$$U^{linear} = \sum_{k=0}^{N-1} \sum_{i=0}^{N-1} P_i \frac{\omega_i^{RF} - \omega_k^{res}(0)(1 + Np_{ki}^{max})}{\Gamma_{kj}^{res}(0)^2(1 + Qp_{ki}^{max})^2 + (\omega_i^{RF} - \omega_k^{res}(0)(1 + Np_{ki}^{max}))^2} \beta, \quad (8)$$

Using for each diode and for each RF signal a single value of the magnetization oscillation power makes the model linear. We chose these values to be the maximums $p_{ki}^{max}$ in order to make the output of the linear model match as much as possible with the output of the realistic model. We repeat the same set of 6561 different calculations that were done with the realistic model (same sweeps of power for the four RF signals and same sweeps of resonance frequencies for each resonator), but this time with the linear model described by Eq. 8. We can then compare the realistic model including non-linearities to a model where the synaptic weights do not depend at all on the input of the synaptic layer.

In Fig. 2.d we plot the voltage of the non-linear MAC simulations as a function of the voltage of the linear MAC. We see that the scatter plot thus created is aligned with the *y=x* curve with a root-mean-square deviation of 0.58 µV. This result shows that the MAC implemented by a chain of spintronic resonators is comparable to a linear MAC when the nonlinear coefficients N and Q are inferior or equal to respectively 0.1 and 1 and thus that spintronic resonators can be used as artificial synapses for neural networks.

## IV. Validation of the model for multiple RF signals superimposition

To simulate the effect of sending simultaneously multiple RF signals in a chain of spintronic resonators, we suppose that the voltage rectified by each resonator is the sum of the voltages it would rectify for each RF signal received individually. This assumption is based on the hypothesis that a spintronic resonator can oscillate simultaneously at different frequencies if it receives different RF signals. To demonstrate this assumption, we make an analysis of the magnetization motion of a spintronic resonator under the influence of different RF signals. We use an Ordinary Differential Equation (ODE) solver to compute the solution of the system of equations of magnetization dynamics [29]:

$$\begin{cases} \frac{dp^{res}}{dt} = -2\Gamma^{res}(p^{res})p^{res} + 2\sqrt{p^{res}} \sum_i^N F_i^{RF} cos(\varphi^{res} - \psi_i^{RF} + \omega_i^{RF} t) \\ \frac{d\varphi^{res}}{dt} = -\omega^{res}(p^{res}) - \frac{1}{\sqrt{p^{res}}} \sum_i^N F_i^{RF} sin(\varphi^{res} - \psi_i^{RF} + \omega_i^{RF} t) \end{cases} \quad (9)$$

where $p^{res}$ and $\varphi^{res}$ are respectively the normalized oscillation power and the phase of the resonator, $\psi_i^{RF}$ the frequencies and phases of incoming RF signals, and $F_i^{RF}$ the amplitude of the torques they exert on the magnetization.

We simulate one spintronic resonator with a frequency $f^{res} = 200$ MHz, a random initial phase $\psi^{res}$ and four RF signals with the same amplitude of microwave torque $F^{RF} = 0.2 \times 2\pi$ rad.MHz, four different frequencies $f_1^{RF} = 200.0$ MHz, $f_1^{RF} = 204.0$ MHz, $f_1^{RF} = 208.2$ MHz, $f_1^{RF} = 212.4$ MHz and four random initial phases $\psi_1^{RF}, \psi_2^{RF}, \psi_3^{RF}$ and $\psi_4^{RF}$. We then compare the horizontal component of the magnetization $m_x(t) = \sqrt{p^{res}(t)} cos(\varphi^{res}(t))$ with our model $m'_x(t) = \sum_i^N \sqrt{p_i^{res}} cos(\psi_i^{RF} + \psi_i^{relaxation} - \omega_i^{RF} t)$. Here each of the normalized oscillation powers $p^{res}(t)$ is calculated using Eq. 7 for a spintronic resonator receiving a single RF signal with microwave power $P_i^{RF}$, frequency $f_i^{RF}$, and initial phase $\psi_i^{RF}$. We considered that the resonator in resonance oscillates at the frequency of the RF signal it receives $f_i^{RF}$. The phases $\psi_i^{relaxation}$ result from the transient dynamics that occurs during the relaxation period (period until the magnetization is periodic), they are determined by fitting the dynamical simulation results to the analytical model. In Fig. 3.a) we see that after this relaxation period, the magnetization dynamics of a spintronic resonator with multiple RF signals corresponds perfectly to our model. Then the resistance oscillations mixed with the RF signals gives $V(\sum_i^N S_i^{RF}) = \sum_i^N S_i^{RF} \times r(t) = \sum_i^N I_i^{RF} cos(\psi_i^{RF} - \omega_i^{RF} t) \times \sum_i^N R_{P-AP} sin(\varphi_i^{res}(t)) \approx \sum_i^N R_{P-AP} I_i^{RF} sin(\varphi_i^{res}(t) - \varphi_i^{RF}(t)) = \sum_i V(S_i^{RF})$, which confirms our assumption. We also repeated all the simulations of the section III with the ODE method. In Fig. 3.b) we compare the voltage of a chain of four resonators simulated with the ODE method and the voltage of the same chain simulated using the analytical model. The voltages are averaged over 20 repetitions of simulations, each time the initial RF signal phases are initialized randomly. The results show that the two models are correlated at 99.94 %. These simulations show that it is valid to consider that the effects of multiple input RF signals simply sum at the resonator level. They validate the use of the analytical model of section III in neural network simulations.

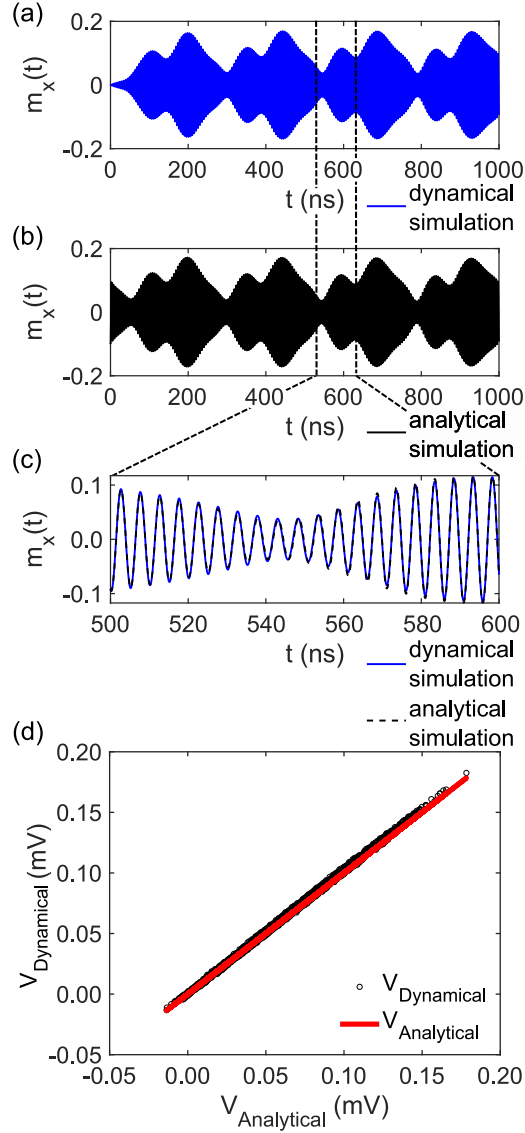

***Fig. 3. (a)*** *Horizontal component of a spintronic resonator magnetization with 4 different radio-frequency signals simulated with an Ordinary Differential Equation solver.* ***(b)*** *Theoretical model* $m'_x(t) = \sum_i^N \sqrt{p_i^{res}} \cos(\psi^{RF} - \omega_i^{RF} t)$. ***(c)*** *ODE simulations (blue solid line) and theoretical model (black dashed line).* ***(d)*** *Simulation of four different radio-frequency signals sent in a chain of four different spintronic resonators for 6561 different combinations of microwave powers for the radio-frequency signals (5 µW, 10 µW, and 15 µW) and different resonance frequencies for the resonators. The scatter dots are the voltages of the ODE simulations plotted against the voltages of the calculations with theoretical model. The simulations are averaged over 20 repetitions, each with a random initialization for the RF signal phases. The red solid line corresponds to the voltages of the simulations realized with the theoretical model plotted against themselves. The root-mean-square deviation between the scatter dots and the red solid line is 1.72 µV and the correlation is 99.94 %*

## V. Handwritten digits recognition with a single layer microwave neural network

In this last section, we prove that we can teach a radio-frequency based neural network to classify microwave-encoded inputs by tuning the resonance frequencies of spintronic resonators, hence demonstrating that the system is able to process RF signals and to directly apply MAC operations on them. To test the efficiency of our implementation of MAC operations for neural networks, we chose a standard task of image classification, for which the goal is to recognize handwritten digits from 0 to 9.

We first consider a dataset called "Digits" of comprising 1797 images of 8 x 8 = 64 pixels. The dataset is split in two: tree quarter of the images are used for the neural network training and one quarter is for testing. The goal for the network is to classify each image between 0 and 9. The network inputs are encoded into 64 RF signals: the brighter the pixel the higher is the RF signal power. The sum of the 64 RF signals is sent into 10 chains of 64 resonators, and the voltages of the 10 synaptic chains are the outputs of the network. The choice of the frequencies of the RF signals, the microwave powers scaling and the initialization of the spintronic resonator frequencies are discussed in appendix A.

To train the network to classify these handwritten digit images we use PyTorch, a software that allows to implement backpropagation, which is the most commonly used algorithm for neural network training [10]. This supervised algorithm propagates the gradient of a loss function across a neural network so that for each iteration, the weight updates $W \leftarrow W - \eta \frac{\partial L}{\partial W}$ reduce the loss $L$, which is the error between the predictions of the network and the targets, i.e. the classification labels assigned to each input. $\eta$ is a learning rate coefficient, that we initialize empirically at $\eta = 10^{-4}$. We use the optimizer Adam [37]. The loss is calculated simulating the voltages $U_j$ of the 10 synaptic chains and applying the Cross Entropy Loss Function

$$L(y, U) = -\sum_{j=0}^{9} y_j \ln\left(\frac{exp(U_j)}{\sum_{j=0}^{9} exp(U_j)}\right) \tag{10}$$

where $y_j$ are the targets.

At each iteration we present a batch of 16 pictures to the network and use Eqs. 2 and 5 with resonator non-linearities to compute the network output. The loss for each picture of the batch is computed and averaged. We then compute the gradient of the loss with respect to the 64x10 weights. To find the updates for the resonance frequencies, using the full nonlinear equations leads to an inefficient backpropagation algorithm because of the dependencies between the synaptic weights and the inputs. However, as the weight changes provoked by backpropagation are by construction small, it is possible to compute them using linearized equations. Therefore, instead of using the model with non-linear resonators, we use the model with linear resonators defined by Eq. 8, initialized with the same parameters as the model with non-linear resonators. To define the reference $p^{max}$, we compute the maximum of magnetization oscillation power for each resonator at initialization for a maximum input (white image, i.e. all the pixels values are one). Then we update the resonance frequencies of the linear model resonators using the weights gradient with respect to the resonance frequencies:

$$f^{res} \leftarrow f^{res} - \eta \frac{\partial W}{\partial f^{res}} \frac{\partial L}{\partial W}. \tag{11}$$

In the next iteration, we take the resonance frequencies that have been updated for the linear model, and use them in the realistic model.

To complete the training procedure, we perform 20 epochs, meaning that we present the entire dataset (training on tree quarter and testing on one quarter) 20 times, and we repeat the entire procedure 10 times to gather statistics. To compute the success rate, i.e., the proportion of images in the dataset that the network is able to classify, we take the class that corresponds to the chain index $j$ whose output is maximum, and we compare it with the target class of the dataset. The purple line in Fig. 4.d shows the mean success rate as a function of the epoch number, and the mean deviation in purple shade. The mean success rate at the end of training reaches 99.96 % both for the test and the training sets. Looking at the standard deviation in purple shade we see that the result is reproducible: if the result is stochastic for the first epochs, the outcome always converges. We performed exactly the same simulations but including the symmetric part for the spin-diode effect with a ratio of 0.5: it means that in Eq. 2 there would be a symmetric part with twice less amplitude than the anti-symmetric part. For this configuration we achieve 99.84 % of success rate. We perform classification on the same task with a classical software neural network trained with backpropagation on an equivalent architecture (64 inputs fully connected by synapses to the 10 outputs). The success rate of the software neural network (blue line in Fig. 4.d) is equivalent to the classification with the resonator network. This result shows that it is possible to train a network made of chains of spintronic resonators by tuning their resonance frequency to classify microwave encoded signals. The training algorithm we developed could also be used to train an experimentally constructed spintronic resonators-based neural network.

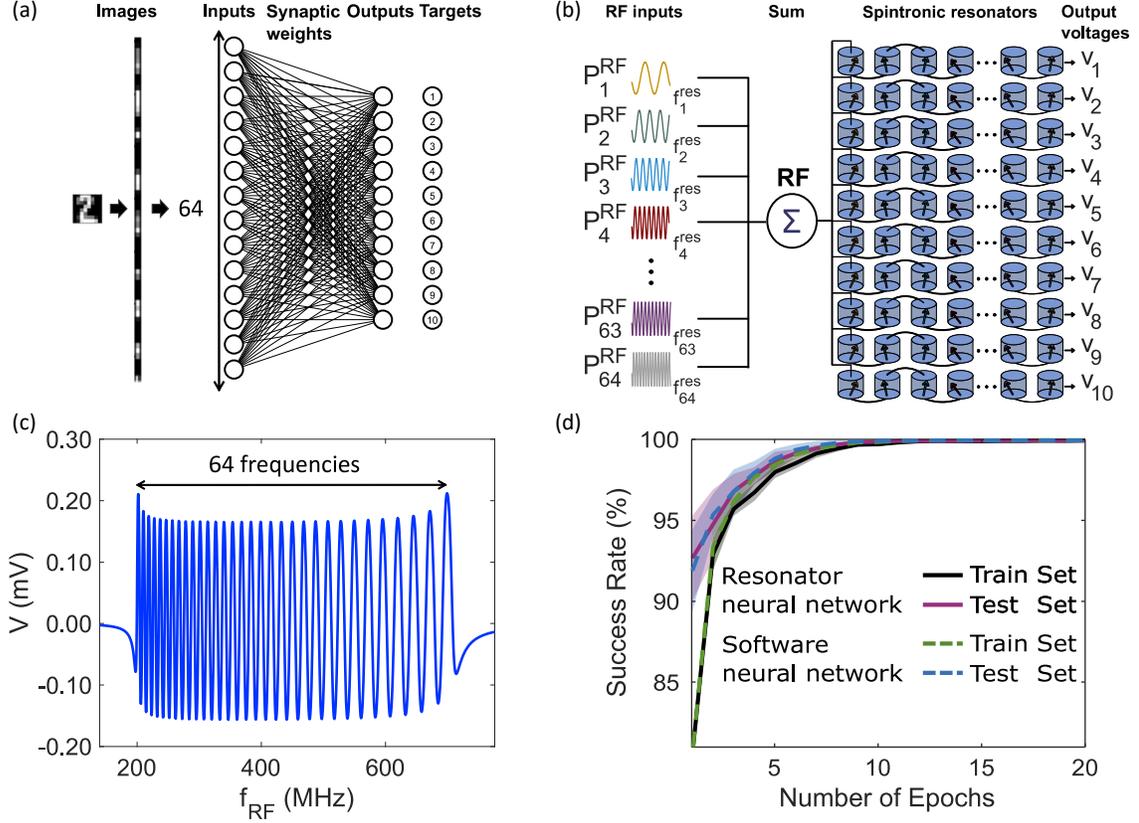

***Fig. 4. (a)*** *Classical neural network architecture to solve the digits dataset. From left to right: 8x8 pixels input images, 64x1 flattened input layer, synaptic layer connecting the input with the 10 outputs, comparison of the outputs with the targets.* ***(b)*** *Equivalent radio-frequency spintronic-synapses-based neural network architecture. From left to right: 8x8 pixels input images, 64x1 flattened input layer, each input is encoded in the microwave power of a radio-frequency signal with a different frequency. The 64 signals are summed and sent to 10 chains of 64 resonators wired in series head-to-tail. Each resonator rectifies its matching frequency signal, thus applying a synaptic weight to it. The output voltages are compared to the targets.* ***(c)*** *Analytical simulations of the spin-diode voltage of a chain of 64 spintronic resonators wired head-to-tail versus the frequency of a RF signal of power 50 µW. The first resonator has a resonance frequency $f_0^{res}$ = 200 MHz and the others are arranged following Eq. 12 of appendix A.* ***(d)*** *Percentage of successful classifications versus number of epochs. Black (purple) color is for the results on the training (test) set for the resonator neural network and green (blue) color is for the train (test) set for the equivalent regular software neural network. The lines (dashed lines for the software neural network) represent the mean success rates and the shade the standard deviations. The success rate reaches 99.96 % both for the software neural network and for the resonator-based neural network for train and test sets.*

Finally, we want to prove that we can scale this type of network to chains comprising several hundreds of spintronic resonators are very large. The goal is to check that the cross-talks due to the superimpositions of many RF signals would not prevent learning. To do so, we solve the "MNIST" dataset which is similar to "digits" but with 70 000 pictures of 28 x 28 = 784 pixels. Then the network is made of 10 chains of 784 spintronic resonators. We use the same algorithm to train this network than the one we used for digits, but with a learning rate $\eta = 5.10^{-6}$ and batches of 500 pictures. Due to the large RAM memory the simulations require, it was not possible to implement the nonlinear behavior of these spintronic resonators for this study.

We arrange the 784 RF frequencies between 50 MHz and 20 GHz as explained Appendix A. For a fixed frequency arrangement, the cross-talks increase with the linewidths of the spintronic resonators, which are proportional to the magnetic damping (see Eq. 2). On the other hand, higher magnetic damping helps to stabilize the orbit of magnetic oscillations faster and leads to higher computational speed. It is possible to engineer the spintronic resonators with different materials to change their magnetic damping.

Fig. 5 we plot the success rate on the "MNIST" dataset with a layer of these spintronic resonators for different magnetic damping after 20 epochs. The results are averaged over 10 repetitions. For the results on the training set (again 1 quarter of the dataset), we found that the maximum recognition rate is 99.40 % for a magnetic damping of α = 0.0188. For this magnetic damping value, the linewidth is comparable to the separation coefficient (see Appendix A): the cross-talks are not dominant and do not prevent the network to learn and to separate the different classes of inputs. In comparison, we solved the same dataset with a software neural network of the same architecture, and achieve at best 92.27 % of recognition. We see that the accuracy of the resonator neural network decreases strongly for α > 0.1. It is because if the magnetic damping is far greater than the separation coefficient, the different resonance curves overlap and the cross-talks degrade classification.

To discuss the relation between this magnetic damping and the computational speed, we have to consider that each iteration is limited by the speed of the slowest spintronic resonator, which in this case is 50 MHz. To estimate the relaxation time of this spintronic resonator, we use an Ordinary Differential Equation solver as we did in section IV to solve Eq. 9. We fit the horizontal component of the magnetization dynamics to an exponential decay model: $m_x^{decay}(t) = \sqrt{p^{res}(t)}\left(1 - e^{-t/\tau}\right)\cos(\varphi^{res}(t))$. Using this technique, we extract that the relaxation time $\tau$ is equal to 194 ns when α = 0.01 (Permalloy) and τ is equal to 20 ns when α = 0.1, which sets the limit speed for the computation.

It is currently difficult to compute the power consumption of such a system, because it would depend strongly on the electronic implementation and on the choice of the spintronic resonators. We can still estimate that the minimum consumption per device due to Joule heating is 1 μW [35]. Since here we use 10 X 784 = 7840 spintronic resonators, the minimum consumption of this resonators neural networks is ~8 mW. It is three orders of magnitude less than the power consumption of a traditional GPU to solve "MNIST" [38].

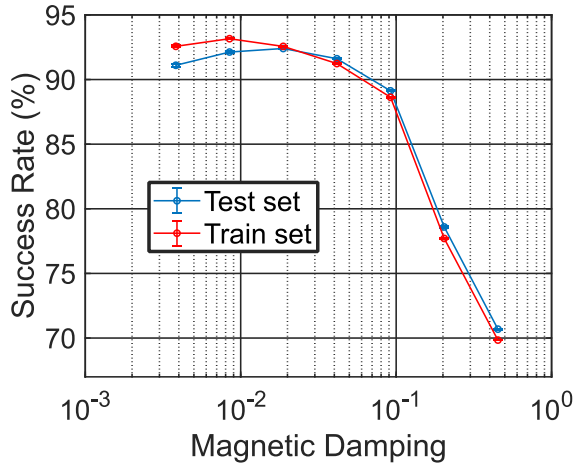

*Fig. 5.* *Percentage of successful classifications of a single layer resonators network on the "MNIST" dataset versus the magnetic damping of the material used for the spintronic resonators (log$_{10}$ scale). In blue (red) are plotted the results for the testing (training) set. The results are averaged over 10 repetitions and the error bar corresponds to the root-mean-square deviation.*

## VI.    Conclusion

To conclude, this work showed theoretically and numerically that it is possible to build a synaptic layer made of chains of spintronic resonators, each resonator emulating a synapse and storing a synaptic weight in its resonance frequency. We demonstrated that the MAC operation thus created is equivalent to a usual software MAC operation and able to classify analog RF signals directly without digitalization. We verified the validity of these results with a realistic model considering the non-linear behaviors of the resonators. We proved that it is possible to train a network of these resonators receiving microwave encoded inputs by changing their resonance frequencies and we achieve software equivalent recognition on the "digits" database. Electric field control of magnetism allows the possibility to have a non-volatile voltage control of these resonance frequencies [32]. Our simulations on the "MNIST" dataset indicate that our network can scale at least to 784 resonators per chain, which shows that this architecture is competitive with memristor crossbar arrays. Building such large array of spintronic resonators will necessitate to optimizing the transmission of RF signals in the resonator chains and addressing impedance mismatches. Our concept using spintronic resonators and frequency-multiplexing provides a fast, compact and low power solution to process Radio-Frequency encoded information with Artificial Intelligence methods.

### Acknowledgments

This work was supported by the European Research Council ERC under Grant No. bioSPINspired 682955, the French ANR project SPIN-IA (Grant No. ANR-18-ASTR-0015) and the French Ministry of Defense (DGA). The authors thank Axel Laborieux and Tifenn Hirtzlin for their scientific support.

## VII.    APPENDIX A: Frequency arrangement and amplification for frequency-multiplexing

To prevent the resonators from rectifying simultaneously several RF signals we have to space their frequencies. They should be arranged in a manner that the whole frequency range is not too wide (spintronic resonators can cover a finite frequency range between few tens of MHz and few tens of GHz [30, 36, 39]) but with resonances overlapping each other as little as possible. We have to consider that for a specific type of spintronic resonators, the higher the frequency is, the wider the linewidth of resonance will be (see Eq. 2). This is because the linewidth scales with the frequency and the magnetic damping α. We want to choose a coefficient of separation that optimizes the spacing between the resonance frequency of the resonators. We define the coefficient of separation μ so that $f_{i+1} - \mu f_{i+1} = f_i + \mu f_i$. This way the spacing between the resonators increases with their linewidth (see Fig. 6).

Hence the RF frequencies follow the law

$$f_i = f_0 \left(\frac{1+\mu}{1-\mu}\right)^i \tag{12}$$

where $f_0$ is the lowest frequency. We chose $f_0 = f_{min} = 50$ MHz to solve the MNIST dataset. To space up the RF frequencies as much as possible, we chose $f_{max} = 20$ GHz in resonance frequency. If the number of RF signals is N, then $f_{N-1} = f_{max}$ and $f_{max} = f_{min}\left(\frac{1+\mu}{1-\mu}\right)^{N-1}$. We can then compute that the optimum spacing coefficient μ is:

$$\mu = \frac{\left(f_{max}/f_{min}\right)^{\frac{1}{N-1}} - 1}{\left(f_{max}/f_{min}\right)^{\frac{1}{N-1}} + 1} \tag{13}$$

In the case of "MNIST" N=784, which gives μ=0.0038. It is important to notice that in eq. 13 the coefficient of separation does not depend directly on the frequency of the resonators but on the ratio between the highest frequency and the lowest one. To solve the dataset "digits", we chose $f_0 = 100$ MHz and μ =α=0.01. In this situation the cross-talks are not dominant because each curve of resonance starts when the previous ends, like it is represented Fig. 6 (μ =α=0.01)

The initialization resonance frequencies of the resonators of each synaptic chain also follow Eq. 12 but with a random shift following a normal distribution with standard deviation $f_k \frac{0.001}{\sqrt{64}}$ for the "digits" dataset, and $f_k \frac{0.001}{\sqrt{784}}$ for "MNIST".

In Eq. 2 of spin-diode voltage we see that the resonator voltage decreases as $\frac{1}{f^{res}}$. Hence in order to obtain comparable signals for all resonators of a chain, we scale the microwave powers of the input layer to increase the signal emitted by the high frequency signals: $P_i \to P_i \frac{f_i^{res}}{f_0^{res}}$.

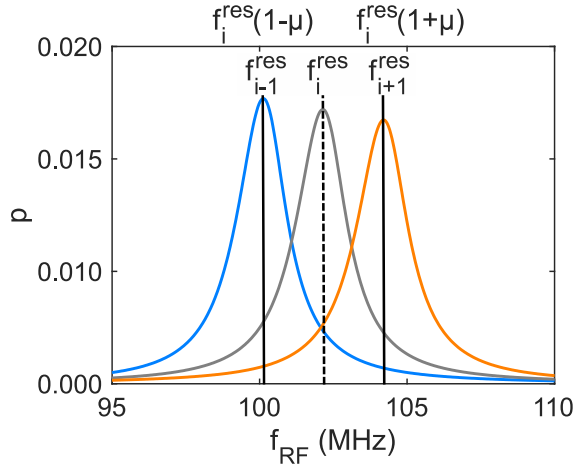

*Fig. 6. Magnetization normalized oscillation power versus frequency of a RF signal for three spintronic resonators of different frequencies following Eq. 12.*

# REFERENCES

[1] T. J. O'Shea, T. Roy, and T. C. Clancy, Over-the-Air Deep Learning Based Radio Signal Classification, IEEE Journal of Selected Topics in Signal Processing 12, 168 (2018)

[2] Y. H. Yoon, S. Khan, J. Huh, and J. C. Ye, Efficient B-Mode Ultrasound Image Reconstruction From Sub-Sampled RF Data Using Deep Learning, IEEE Transactions on Medical Imaging 38, 325 (2019)

[3] M. Dai, S. Li, Y. Wang, Q. Zhang, and J. Yu, Post-Processing Radio-Frequency Signal Based on Deep Learning Method for Ultrasonic Microbubble Imaging, BioMedical Engineering OnLine 18, 95 (2019)

[4] E. Besler, Y. C. Wang, and A. V. Sahakian, Real-Time Radiofrequency Ablation Lesion Depth Estimation Using Multi-Frequency Impedance With a Deep Neural Network and Tree-Based Ensembles, IEEE Transactions on Biomedical Engineering 67, 1890 (2020)

[5] K. Merchant, S. Revay, G. Stantchev, and B. Nousain, Deep Learning for RF Device Fingerprinting in Cognitive Communication Networks, IEEE Journal of Selected Topics in Signal Processing 12, 160 (2018)

[6] J. Lien, N. Gillian, M. E. Karagozler, P. Amihood, C. Schwesig, E. Olson, H. Raja, and I. Poupyrev, Soli: Ubiquitous Gesture Sensing with Millimeter Wave Radar, ACM Trans. Graph. 35, 142 (2016)

[7] Y. Kim, Application of Machine Learning to Antenna Design and Radar Signal Processing: A Review, in 2018 International Symposium on Antennas and Propagation, 1–2 (2018)

[8] M. F. Al-Sa'd, A. Al-Ali, A. Mohamed, T. Khattab, and A. Erbad, RF-Based Drone Detection and Identification Using Deep Learning Approaches: An Initiative towards a Large Open Source Drone Database, Future Generation Computer Systems 100, 86 (2019)

[9] E. García-Martín, C. F. Rodrigues, G. Riley, and H. Grahn, Estimation of Energy Consumption in Machine Learning, Journal of Parallel and Distributed Computing 134, 75 (2019)

[10] Y. LeCun, Y. Bengio, and G. Hinton, Deep learning, Nature, 521, 436–444, (2015)

[11] Big data needs a hardware revolution, Nature 554, 145–146 (2018)

[12] Schuman, C. D, T. E. Potok, R. M. Patton, J. D. Birdwell, M. E. Dean, G. S. Rose, J. S. Plank, A survey of neuromorphic computing and neural networks in hardware, arXiv 1705.06963 (2017)

[13] S. Ambrogio, P. Narayanan, H. Tsai, R. M. Shelby, I. Boybat, C. di Nolfo, M. Giordano, M. Bodini, N. C. P. Farinha, B. Killeen, C. Cheng, Y. Jaoudi & G. W. Burr, Equivalent-accuracy accelerated neural-network training using analogue memory, Nature 558, 60–67 (2018)

[14] F. Cai, J. M. Correll, S. H. Lee, Y. Lim, V. Bothra, Z. Zhang, M. P. Flynn and W. D. Lu, A fully integrated reprogrammable memristor–CMOS system for efficient multiply–accumulate operations, Nature Electronics 2, 290–299 (2019)

[15] P. Yao, H. Wu, B. Gao, J. Tang, Q. Zhang, W. Zhang, J. Joshua Yang and H. Qian, Fully hardware-implemented memristor convolutional neural network, Nature 577, 641–646 (2020)

[16] R. Hamerly, L. Bernstein, A. Sludds, M. Soljačić, and D. Englund, Large-Scale Optical Neural Networks Based on Photoelectric Multiplication, Phys. Rev. X, 9, 021032 (2019)

[17] J. Feldmann, N. Youngblood, M. Karpov, H. Gehring, X. Li, M. Stappers, M. Le Gallo, X. Fu, A. Lukashchuk, A. S. Raja, J. Liu, C. D. Wright, A. Sebastian, T. J. Kippenberg, W. H. P. Pernice and H. Bhaskaran, Parallel convolution processing using an integrated photonic tensor core, Nature 589, 82–58 (2021)

[18] J. Torrejon, M. Riou, F. Abreu Araujo, S. Tsunegi, G. Khalsa, D. Querlioz, P. Bortolotti, V. Cros, K. Yakushiji, A. Fukushima, H. Kubota, S. Yuasa, M. D. Stiles and J. Grollier, Neuromorphic computing with nanoscale spintronic oscillators, Nature 547, 428 (2017)

[19] D. Marković, N. Leroux, M. Riou, F. Abreu Araujo, J. Torrejon, D. Querlioz, A. Fukushima, S. Yuasa, J. Trastoy, P. Bortolotti, and J. Grollier, Reservoir Computing with the Frequency, Phase, and Amplitude of Spin-Torque Nano-Oscillators, Appl. Phys. Lett. 114, 012409 (2019)

[20] S. Tsunegi, T. Taniguchi, K. Nakajima, S. Miwa, K. Yakushiji, A. Fukushima, S. Yuasa, and H. Kubota, Physical Reservoir Computing Based on Spin Torque Oscillator with Forced Synchronization, Appl. Phys. Lett. 114, 164101 (2019)

[21] M. Zahedinejad, A. A. Awad, S. Muralidhar, R. Khymyn, H. Fulara, H. Mazraati, M. Dvornik, and J. Åkerman, Two-Dimensional Mutually Synchronized Spin Hall Nano-Oscillator Arrays for Neuromorphic Computing, Nat. Nanotechnol, 15, 47 (2020)

[22] M. Koo, M. R. Pufall, Y. Shim, A. B. Kos, G. Csaba, W. Porod, W. H. Rippard, and K. Roy, Distance Computation Based on Coupled Spin-Torque Oscillators: Application to Image Processing, Phys. Rev. Applied 14, 034001 (2020)

[23] H. Arai and H. Imamura, Neural-Network Computation Using Spin-Wave-Coupled Spin-Torque Oscillators, Phys. Rev. Applied 10, 024040 (2018)

[24] D. E. Nikonov, P. Kurahashi, J. S. Ayers, H.-J. Lee, Y. Fan, and I. A. Young, Convolution Inference via Synchronization of a Coupled CMOS Oscillator Array, IEEE Journal on Exploratory Solid-State Computational Devices and Circuits 6, 170–176 (2020)

[25] D. E. Nikonov, G. Csaba, W. Porod, T. Shibata, D. Voils, D. Hammerstrom, I. A. Young, and G. I. Bourianoff, Coupled-Oscillator Associative Memory Array Operation for Pattern Recognition, IEEE Journal on Exploratory Solid-State Computational Devices and Circuits 1, 85 (2015)

[26] A. A. Tulapurkar, Y. Suzuki, A. Fukushima, H. Kubota, H.Maehara, K. Tsunekawa, D. D. Djayaprawira, N.Watanabe, and S. Yuasa, Spin-torque diode effect in magnetic tunnel junctions, Nature 438, 339 (2005)

[27] B. Fang, M. Carpentieri, X. Hao, H. Jiang, J. A. Katine, I. N. Krivorotov, B. Ocker, J. Langer, K. L. Wang, B. Zhang, B. Azzerboni, P. Khalili Amiri, G. Finocchio and Z. Zeng, Giant spin-torque diode sensitivity in the absence of bias magnetic field, Nat. Commun 7, 11259 (2016)

[28] J. Cai, L. Zhang, B. Fang, W. Lv, B. Zhang, G. Finocchio, R. Xiong, S. Liang, and Z. Zeng, Sparse Neuromorphic Computing Based on Spin-Torque Diodes, Applied Physics Letters 114, 192402 (2019)

[29] A. Slavin and V. Tiberkevich, Nonlinear Auto-Oscillator Theory of Microwave Generation by Spin-Polarized Current, IEEE Transactions on Magnetics, 45, 1875–1918, (2009)

[30] A. Dussaux, A. V. Khvalkovskiy, P. Bortolotti, J. Grollier, V. Cros, and A. Fert, Field dependence of spin-transfer-induced vortex dynamics in the nonlinear regime, Phys. Rev. B 86, 014402 (2012)

[31] A. Litvinenko, V. Iurchuk, P. Sethi, S. Louis, V. Tyberkevych, J. Li, A. Jenkins, R. Ferreira, B. Dieny, A.Slavin, U. Ebels, Ultrafast Sweep-Tuned Spectrum Analyzer with Temporal Resolution Based on a Spin-Torque Nano-Oscillator, Nano Lett. 20, 8, 6104–6111 (2020)

[32] M. Zahedinejad, H. Fulara, R. Khymyn, A. Houshang, S. Fukami, S. Kanai, H. Ohno, and J. Åkerman, Memristive Control of Mutual SHNO Synchronization for Neuromorphic Computing, ArXiv:2009.06594 [Physics] (2020)

[33] D. Marković, N. Leroux, A. Mizrahi, J. Trastoy, V. Cros, P. Bortolotti, L. Martins, A. Jenkins, R. Ferreira, and J. Grollier, Detection of the Microwave Emission from a Spin-Torque Oscillator by a Spin Diode, Phys. Rev. Applied 13, 044050 (2020)

[34] S. Jiang, R. Khymyn, S. Chung, T. Quang Le, L. H. Diez, A. Houshang, M. Zahedinejad, D. Ravelosona, and J. Åkerman, Reduced spin torque nano-oscillator linewidth using He+ irradiation Appl. Phys. Lett. 116, 072403 (2020)

[35] M. Romera, P. Talatchian, S. Tsunegi, F. Abreu Araujo, V. Cros, P. Bortolotti, J. Trastoy, K. Yakushiji, A. Fukushima, H. Kubota, S. Yuasa, M. Ernoult, D. Vodenicarevic, T. Hirtzlin, N. Locatelli, D. Querlioz and J. Grollier, Vowel recognition with four coupled spin-torque nano-oscillators, Nature, 563, 230–234 (2018)

[36] B. Divinskiy, S. Urazhdin, S. O. Demokritov and V. E. Demidov, Controlled nonlinear magnetic damping in spin-Hall nano-devices, Nat. Commun 10, 5211 (2019)

[37] D. P. Kingma and J. Ba, Adam: A Method for Stochastic Optimization, ArXiv:1412.6980 (2014)

[38] V. Joseph, C. Nagarajan, MADONNA: A Framework for Energy Measurements and Assistance in designing Low Power Deep Neural Networks

[39] S. Bonnetti, P. Muduli, F. Mancoff and J. Åkerman, Spin torque oscillator frequency versus magnetic field angle: The prospect of operation beyond 65 GHz, Appl. Phys. Lett. 94, 102507 (2009).